\documentclass[intlimits,twoside,a4paper]{article}

\usepackage{amsmath,amssymb} 
\usepackage{graphicx}        

\usepackage[utf8]{inputenc}

\usepackage[eqsecnum]{cmpj3}

\usepackage[export]{adjustbox}

\newcommand{\be}{\begin{equation}}
\newcommand{\ee}{\end{equation}}

\issue{2026}{29}{1}{13403}
\doinumber{10.5488/CMP.29.13403}

\title[Antibody fluid constrained by rigid spherical obstacles]%
{Behaviour of the model antibody fluid constrained by rigid spherical obstacles: effects of the obstacle-antibody binding}
\author[Yu. V. Kalyuzhnyi, T.~Patsahan]{Yu. V. Kalyuzhnyi\orcid{0000-0002-0631-9982}\refaddr{label1,label2}\thanks{Corresponding author: \email{yukal@icmp.lviv.ua}},
	T.~Patsahan\orcid{0000-0002-7870-2219}\refaddr{label2,label3}}
\addresses{
	\addr{label1} University of Ljubljana, Faculty of Chemistry and Chemical Technology,  Ve\v{c}na pot 113, SI-1000
Ljubljana, Slovenia
    \addr{label2} Yukhnovskii Institute for Condensed Matter Physics of the National Academy of Sciences of Ukraine, 1 Svientsitskii Str., 79011 Lviv, Ukraine
	\addr{label3} Institute of Applied Mathematics and Fundamental Sciences, Lviv Polytechnic National University, 12~S.~Bandera~Str., 79013 Lviv, Ukraine}

\date{Received 19 February 2026; revised 4 March 2026; accepted 4 March 2026; published 30 March 2026}
\begin{document}
	
	\maketitle
	
	\begin{abstract}

We study a simplified model of monoclonal antibodies confined in a patchy random porous medium. Antibodies are represented as Y-shaped particles composed of seven tangential hard spheres with attractive patches on the terminal beads, while the matrix consists of randomly distributed hard-sphere obstacles bearing adhesive sites. The model captures antibody behavior in crowded biological environments with strong short-range antibody-matrix attractions. The theoretical approach combines Wertheim’s multidensity thermodynamic perturbation theory, the Flory-Stockmayer theory of polymerization, and scaled particle theory for fluids in porous media. We analyze thermodynamic properties, percolation thresholds, and phase behavior, and compare the selected results with new computer simulations. The interplay between antibody-antibody and antibody-matrix interactions produces a complex phase behavior, including re-entrant phase separation with a closed-loop coexistence region at higher temperatures and conventional liquid-gas separation at lower temperatures.
    
\printkeywords  
	\end{abstract}
	
\section{Introduction}

Over the past few decades, considerable efforts have been devoted to understanding how the crowded and confined environment of the living cells affects the properties of proteins 
{
(see \cite{das2024macromolecular} and references therein).}
The intracellular environment is highly crowded with a wide variety of macromolecular species, including nucleic acids, ribosomes, other proteins, electrolytes, etc. Due to excluded-volume interactions and the associated reduction of free volume (these species are commonly referred to as crowders) restrict the translational and rotational motion of protein molecules. These excluded volume effects, which have been extensively studied, constitute the so-called hard (repulsive) component of macromolecular crowding.
A second contribution, arising from { additional}
soft (attractive) interactions between proteins and their surrounding medium, has received comparatively less attention and will be partially examined in the present work.

Recently, a coarse-grained model for aqueous solutions of monoclonal antibodies (mAbs) has been proposed~\cite{kalyuzhnyi2018modeling,kastelic2018controlling,kastelic2018theory,vlachy2023protein}. In this model, mAb macromolecules are represented as assemblies of seven hard-sphere beads arranged to form a Y-shaped structure, with additional attractive patches located on the terminal beads. This model has been successfully used to analyze experimental data on viscosity~\cite{kastelic2018controlling,hvozd2024modelling} and phase behavior~\cite{kalyuzhnyi2018modeling,kastelic2018theory,kalyuzhnyi2022numerical,hvozd2024modelling} 
of antibody solutions in aqueous electrolytes.
In a more recent work, we employed the seven-bead antibody model to investigate the effects of the cellular environment on antibody solutions, with particular emphasis on the phase behavior and percolation properties~\cite{hvozd2020aggregation,hvozd2022behaviour,Hvozd2022}.
To describe crowding and confinement, the biological milieu of the cell was mimicked by a collection of immobile hard-sphere obstacles randomly distributed in space, within which antibody macromolecules are free to move. These obstacles represent slowly moving or effectively immobile crowders. Such systems can be viewed as partly quenched~\cite{hribar2011partly}, with protein degrees of freedom annealed and crowder degrees of freedom quenched. In two preceding studies, we considered models with inert (purely repulsive, hard) obstacles~\cite{hvozd2020aggregation} and with obstacles interacting attractively via a weak Yukawa potential (soft component)~\cite{hvozd2022behaviour}. The excluded-volume interactions were shown to reduce the critical temperature and density, and to broaden the percolation region~\cite{hvozd2020aggregation}. Even more intriguing behavior was observed when weak attractive interactions between antibodies and obstacles were introduced: the resulting phase diagram exhibited re-entrant behavior, with the system passing from a one-phase region to a two-phase region and back to a one-phase region upon cooling. Simultaneously, the binodal curve narrowed and the system approached the so-called empty liquid regime~\cite{Hvozd2022}, where the liquid branch of the phase diagram shifts toward very low densities.

The present work focuses on the effects of strong binding interactions between antibody macromolecules and matrix obstacles on the phase behavior, aggregation, and percolation properties of the system. To this end, we consider a model consisting of seven-bead antibody macromolecules confined within a matrix of randomly distributed hard-sphere obstacles, each decorated with a certain number of sticky spots (patches) capable of forming bonds with the corresponding patches on the antibodies. The theoretical description combines thermodynamic perturbation theory (TPT) for associating fluids, the Flory-Stockmayer theory (FST) of polymerization, and an extension of scaled particle theory (SPT) for a fluid in porous media~\cite{Wertheim3,Wertheim4,Wertheim1987,holovko2012fluids,patsahan2011fluids,holovko2017improvement,kalyuzhnyi2014phase,bianchi2008theoretical,bianchi2007fully,de2011phase,tavares2010equilibrium}.

Our study was motivated in part by recent work of Kalyuzhnyi et al.~\cite{kalyuzhnyi2024phase}, where highly nontrivial phase behavior was reported for patchy colloids confined in a matrix of patchy obstacles. In that system, competition between inter-colloidal bonding and bonding between colloidal particles and matrix obstacles led to a re-entrant phase behavior featuring three critical points and two distinct liquid-gas coexistence regions. In the present study, we extend these ideas by explicitly accounting for the nonspherical shape of antibody macromolecules and their flexibility, which play a crucial role in determining the collective behavior of realistic protein systems.

This paper is dedicated to the 75th anniversary of the birth of our late friend and colleague, Stefan Soko{\l}owski, who made important contributions to the statistical-mechanical theory of fluid adsorption and interfacial phenomena. Among his notable achievements is the development of the replica Ornstein-Zernike theory for associative fluids and hard-sphere systems confined in random and polydisperse porous media~\cite{trokhymchuk1996adsorption,trokhymchuk1997associative,ilnytsky1999replica,pizio2000effects,rzysko2002theory}.

\section{The model and theory}

\begin{figure}[!htb] 	
	\begin{center}
		\includegraphics[width = 0.15\textwidth,  valign=c]{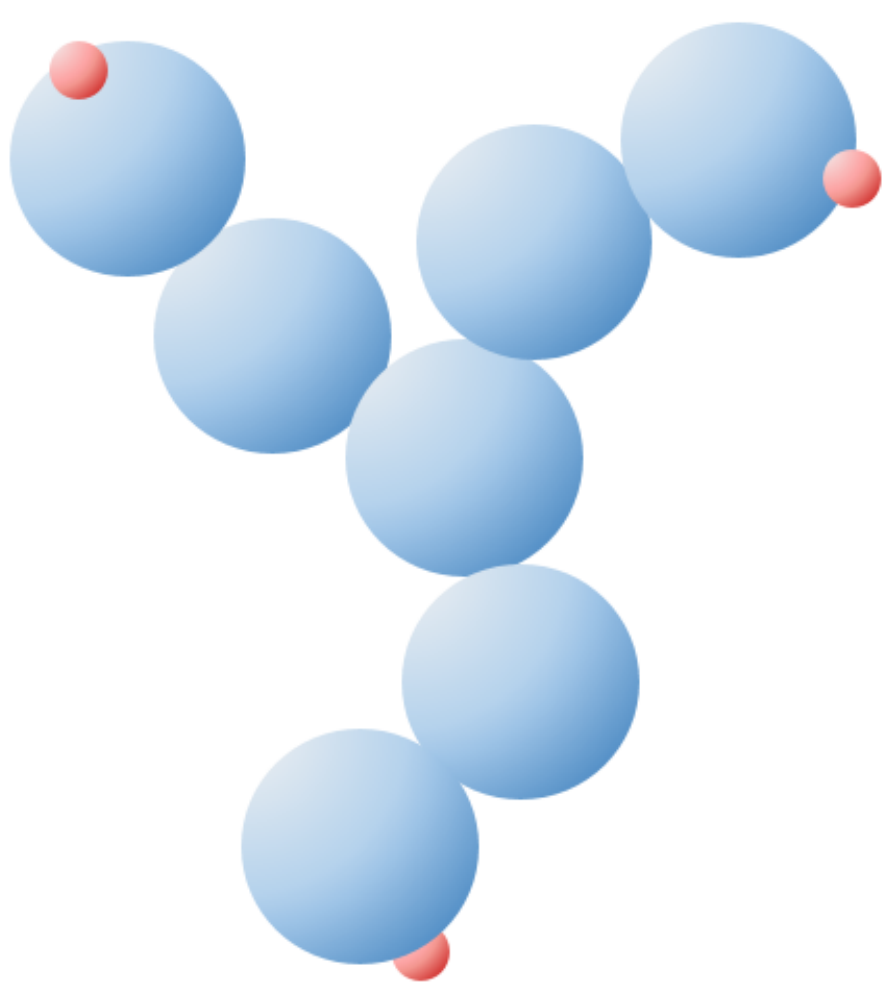} \qquad   
		\includegraphics[width = 0.45\textwidth,  valign=c]{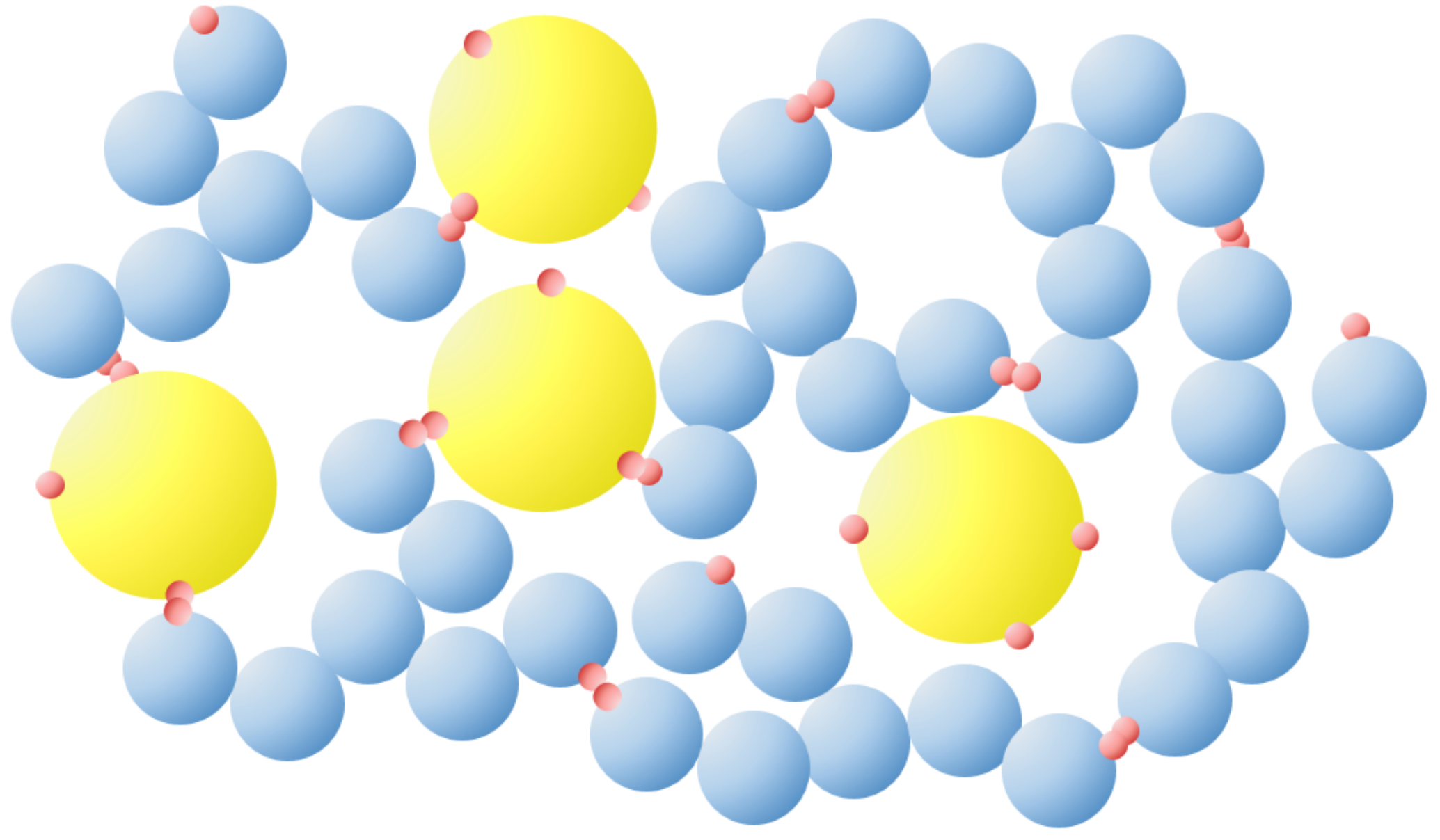}
		\caption{(Colour online) Coarse-grained model of antibody molecules (blue) confined within a matrix of rigid spherical obstacles (yellow), both decorated with attractive patches (red). 
			{The left-hand panel shows an example of an individual antibody molecule built from seven tangentially connected beads.}}
		\label{model}
	\end{center}
\end{figure}

We consider a solution of antibody molecules modelled by the collection of seven hard spheres 
of the size $\sigma_1$, tangentially bonded to form three-arm completely flexible symmetrical Y-shaped object  (figure~\ref{model}). 
Each of three terminal hard-sphere beads is decorated by off-center square-well site located on its surface. 
The fluid of antibody molecules is confined in the matrix of patchy hard-sphere obstacles of the size $\sigma_0$, 
randomly distributed in space and decorated by several off-center square-well sites. Interaction between the 
terminal beads and between terminal beads and obstacles is represented by the following pair potential:
\be
U_{i_Kj_L}(12)=U^{(\rm{hs})}_{ij}(r)+U_{i_Kj_L}^{(\rm{as})}(12),
\label{UKL}
\ee
where
\be
U_{i_Kj_L}^{(\rm{as})}(12)=\left\{
\begin{array}{rl}
	-\epsilon_{ij}, & {\rm for}\;z(12)\leqslant\omega_{ij}\\
	0, & {\rm otherwise}
\end{array}
\right.,
\label{Uass} 
\ee 
$1(2)$ denote position and orientation of the particle $1(2)$, the indices $i,j$ take the values
$(i,j)=(1,1),(1,0),(0,1)$ and denote either obstacles ($i=0$) or antibody molecules ($i=1$)
$U^{(\rm{hs})}_{ij}(r)$ is the hard-sphere potential, $K$ and $L$ denote the sites and take the values $A,B,C\ldots$,  
$\omega_{ij}$ and $\epsilon_{ij}\;(>0)$ are the width and depth of the off-center square-well sites, respectively,
and $z(12)$ is the distance between these square-well sites. Note that due to expression (\ref{Uass}) for 
$U_{i_Kj_L}^{(\rm{as})}(12)$ the sites which belong to
the particles of the same type are equivalent. The number densities of the matrix and molecules are $\rho_0$ and $\rho_1$, respectively, and the temperature of the system is $T$.

Thermodynamics of the model at hand is calculated using an appropriate combination of the SPT and TPT. In the
framework of the TPT we have:
\be
A=A_{\rm{ref}}+\Delta A_{\rm{as}},
\label{ATPT}
\ee
where $A$ is Helmholtz free energy of the system, $A_{\rm{ref}}$ is Helmholtz free energy of the reference system 
and $\Delta A_{\rm{as}}$ is the corresponding contribution due to associative interaction $U_{i_Kj_L}^{(\rm{as})}(12)$.
The reference system is represented by the original system with $\epsilon_{ij}=0$, i.e., by the fluid of
non-associating molecules confined in the matrix of the hard-sphere obstacles. We have
\be
{\beta A_{\rm{ref}}\over V}=\rho_1\ln{\left(\Lambda^3\rho_1\right)}-\rho_1-6\rho_1\ln{\left(g^{(\rm{hs})}_{11}\right)}+
{\beta\Delta A_{\rm{hs}}\over V},
\label{Aref}
\ee
where $\beta=1/k_{\rm{B}}T$, $k_{\rm{B}}$ is the Boltzmann constant,
$\Lambda$ is de Broglie thermal wavelength, 
$g_{11}^{(\rm{hs})}=g_{11}^{(\rm{hs})}(r=\sigma_{1})$, $g_{11}^{(\rm{hs})}(r)$ and 
$\Delta A_{\rm{hs}}$ are radial distribution function (RDF) and excess Helmholtz free energy of hard-sphere 
fluid confined in the hard-sphere matrix, respectively. Here the number density of hard-sphere 
fluid is $\rho_1^{(\rm{hs})}=7\rho_1$. The properties of the reference system can be calculated analytically 
using SPT approach~\cite{holovko2012fluids,patsahan2011fluids,holovko2017improvement,kalyuzhnyi2014phase}. 
According to TPT for associating fluids the expression for $\Delta A_{\rm{as}}$ is
\be
\frac{\beta\Delta A_{\rm{as}}}{V}=3\rho_1\left(\ln{X_1}-\frac{1}{2}X_1+\frac{1}{2}\right),
\label{Aas}
\ee
where $X_i$ is fraction of the particles of the type $i$ with one certain site non-bonded and $n_0$
is the number of sites on the matrix obstacles. These fractions satisfy the following equation
\cite{kalyuzhnyi2024phase,Wertheim3,Wertheim1987}:
\be
12\piup \rho_1\sigma_{11}^3X_1^2B_{11}^{(\rm{as})}g_{11}^{(\rm{ref})}+
{4\piup X_1\rho_0\sigma_{10}^3n_0B_{10}^{(\rm{as})}g_{10}^{(\rm{ref})}\over 1+12\piup \rho_1\sigma_{10}^3X_1B_{10}^{(\rm{as})}g_{10}^{(\rm{ref})}}+X_1-1=0,
\label{X1}
\ee
where
$\sigma_{ij}=(\sigma_i+\sigma_j)/2$, $g_{11}^{(\rm{ref})}$ and $g_{10}^{(\rm{ref})}$ are the contact values of 
the site-site RDF between the terminal monomers of antibody molecules and between the terminal monomers
of antibody molecules and obstacles of the matrix, respectively,
\be
\sigma_{ij}^3B_{ij}^{(\rm{as})}=\int_{\sigma_{ij}}^{\sigma_{ij}+\omega_{ij}}{\tilde f}_{ij}^{(\rm{as})}(r)r^2\rd r,
\label{Bij}
\ee
${\tilde f}_{ij}^{(\rm{as})}(r)$ is an orientation averaged Mayer function for off-center site-site square-well 
interaction acting between two hard spheres of the type $i$ and $j$, i.e.,
\be
{\tilde f}^{(\rm{as})}_{ij}(r)=
\left(\re^{\beta\epsilon_{ij}}-1\right)\left(\omega_{ij}+\sigma_{ij}-r\right)^2
\left(2\omega_{ij}-\sigma_{ij}+r\right)/\left(6\sigma_i\sigma_j r\right).
\label{Mayer}
\ee
Here~\cite{kalyuzhnyi1995solution,kalyuzhnyi1997primitive,lin1998j,butovych2023modeling}
\be
g_{11}^{(\rm{ref})}=g_{11}^{(\rm{hs})}-{1\over 2(1-\eta)},\;\;\;\;
g_{10}^{(\rm{ref})}=g_{10}^{(\rm{hs})}-{1\over 4(1-\eta)},
\label{g11g10}
\ee
\noindent
where $\eta=\eta_0+\eta_1$, $\eta_0=\piup\rho_0\sigma_0^3/6$, $\eta_1=\pi\rho_1\sigma_1^3/6$,
$g_{10}^{(\rm{hs})}$ is the contact values of the radial distribution function between the particles of hard-sphere fluid with the number density $\rho_1^{(\rm{hs})}$ 
and obstacles of the matrix.

Finally for chemical potential $\mu$ and pressure $P$ we have
\be
\mu=\mu_{\rm{ref}}+\Delta\mu_{\rm{as}},\;\;\;\;P=P_{\rm{ref}}+\Delta P_{\rm{as}},
\label{finally}
\ee
where 
\be
\Delta\mu_{\rm{as}}=\left({\partial(\Delta A_{\rm{as}}/V)\over\partial\rho_1}\right)_{T,V},\;\;
\Delta P_{\rm{as}}=\rho_1\Delta\mu_{\rm{as}}-\Delta A_{\rm{as}}/V.
\label{standart}
\ee
These expressions are used to calculate liquid-liquid phase diagram of the system using the solution of the 
following set of two equations:
\be
\left\{
\begin{array}{rl}
	\mu(T,\rho_1^{(g)})= & \mu(T,\rho_1^{(\rm{\ell})})\\
	P(T,\rho_1^{(g)})= & P(T,\rho_1^{(\rm{\ell})})
\end{array}
\right.,
\label{phase} 
\ee 
which represent phase equilibrium conditions. Here $\rho_1^{(g)}$ and $\rho_1^{(\ell)}$ are coexisting densities
of the ``gas'' (low density) and ``liquid'' (high density) phases, respectively. 
{ This set of equations was solved numerically using standard methods.}
Our results for the 
fractions $X_0$ and $X_1$ are used to calculate the percolation threshold line. According to the extended 
version of FS theory \cite{kalyuzhnyi2024phase} percolation threshold line for our model satisfies the following equality:
\be
p_{11}(\rho_1,T)=1/2,
\label{percolation}
\ee
where $p_{11}$ is the probability of the molecules to form a bond, i.e.,
\be
p_{11}(\rho_1,T)=1-X_1-{\rho_0\over 3\rho_1}(1-X_0).
\label{p11}
\ee

\section{Computer simulation details}

Computer simulations of the coarse-grained model of antibody molecules confined in a matrix of patchy obstacles are performed using an approach similar to that applied in~\cite{butovych2023modeling}, where a mixture of chain molecules and monomers decorated with attractive patches were studied. 
Within this framework, the hard-sphere potential and square-well patch-patch interaction are replaced by their continuous analogues, namely the pseudo-hard-sphere (PHS) potential~\cite{jover2012pseudo} and continuous square-well (CSW) potential~\cite{espinosa2019breakdown}, which makes the model suitable for molecular dynamics (MD) simulations. Such patchy molecular chain models have been successfully applied to investigate the connectivity and phase behaviour in biomolecular systems, including biomolecular condensates and protein-RNA mixtures~\cite{espinosa2020liquid,joseph2021thermodynamics,sanchez2022rna}.
In the present study we follow the same scheme and simulate the model described in the previous section (figure~\ref{model}) by considering a mixture
of antibody Y-shaped molecules and patchy obstacles, where the degrees of freedom of obstacles are completely frozen.

The MD simulations were carried out in the $NVT$ ensemble using the LAMMPS simulation package (version 29 Aug 2024, Update 2)~\cite{plimpton1995fast, thompson2022lammps}. The temperature was maintained using a Langevin thermostat~\cite{allen2017computer}, which couples each particle to an implicit heat bath through dissipative (frictional) and stochastic (thermal noise) forces.
{The CSW potential $u_{\mathrm{CSW}}(r)$ was taken in the form~\cite{espinosa2019breakdown} 
\begin{equation}
u_{\mathrm{CSW}}(r) = -\frac{1}{2}\,\epsilon_{ij}
\left[
1 - \tanh\left( \frac{r - r_w}{\alpha} \right)
\right],
\end{equation}
while the parameter $\alpha = 0.001\sigma$ was employed, the attractive well width was set to $r_w = 0.12\sigma$ and the cutoff radius to $0.14\sigma$.} 
With this parameterization, the shape of $u_{\mathrm{CSW}}(r)$ closely reproduces that of a conventional square-well potential with depth $\epsilon_{ij}$ and width $r_w$.
The positions of patches were constrained to the surfaces of host beads by fixing their distance from the corresponding bead centers $\sigma_{ij}/2$ using the SHAKE algorithm, with a tolerance parameter of $10^{-4}$. The diameters of beads forming the antibody molecules and obstacles were set to $\sigma_1 = \sigma$ and $\sigma_0 = 3\sigma$, respectively, where $\sigma$ defines the unit of length.
{The pseudo-hard-sphere (PHS) potential describing the interactions between beads of molecules, as well as between beads of molecules and obstacles, has the form of the repulsive part of the cut and shifted (50,49)-Mie potential:
\begin{equation}
u_{ij}^{\mathrm{PHS}}(r) =
\begin{cases}
50 \left( \frac{50}{49} \right)^{49} \epsilon_{\mathrm{PHS}}
\left[
\left( \dfrac{\sigma_{ij}}{r} \right)^{50}
-
\left( \dfrac{\sigma_{ij}}{r} \right)^{49}
\right],
& r < \frac{50}{49}\,\sigma_{ij}, \\[10pt]
0,
& r \geqslant \frac{50}{49}\,\sigma_{ij}.
\end{cases}
\end{equation}
The parameters were employed as suggested in~\cite{jover2012pseudo}. Specifically, $\epsilon_{\mathrm{PHS}} = \epsilon T^*/1.5$, where $T^* = k_{\text{B}} T / \epsilon$ denotes the reduced temperature, and $\epsilon$ corresponds to the energy units,
while $\sigma_{11}=\sigma_{1}$ and $\sigma_{01}=(\sigma_{0}+\sigma_{1})/2$.}
Intramolecular connectivity within the Y-shaped antibody molecules was modelled by harmonic bonds, $u_{\mathrm{bond}}(r) = K (r - r_0)^2$ {applied between centers of adjacent beads},
with a spring constant $K = 160.0\,\epsilon/\sigma^2$ and an equilibrium bond length $r_0 = \sigma_1$. 
{Accordingly, a fully flexible seven-bead molecule was composed of a central bead 
to which three arms were attached, each consisting of two beads (see figure~\ref{model}).}
PHS interactions between adjacent beads in the same molecule were excluded. In addition, CSW interactions between patches belonging to the same molecule were suppressed, thus intramolecular patch-patch binding was not permitted, consistent with the assumptions of the theory presented in this study.

\begin{figure}[!htb] 	
	\begin{center}
		\includegraphics[width = 0.45\textwidth]{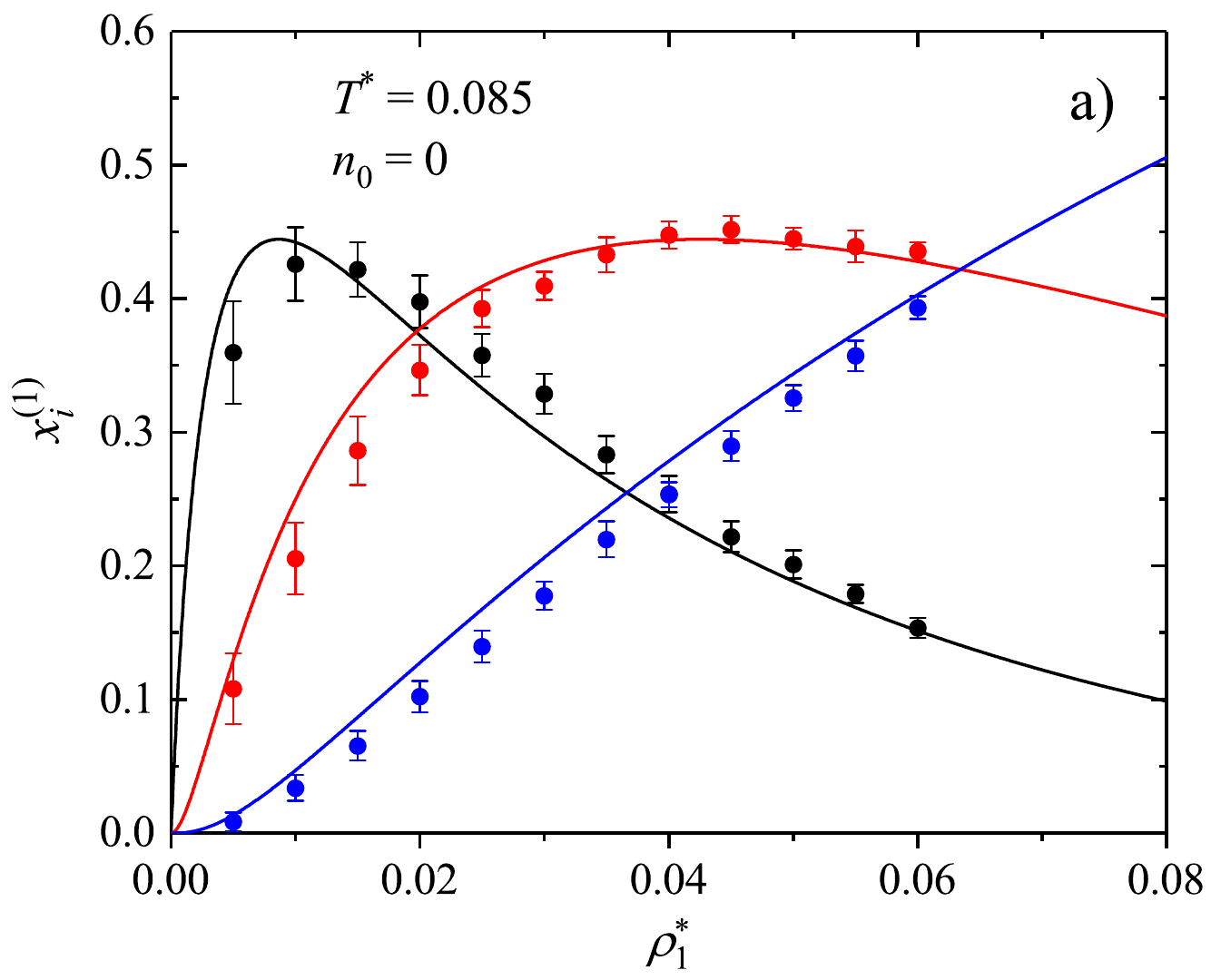}
		\includegraphics[width = 0.45\textwidth]{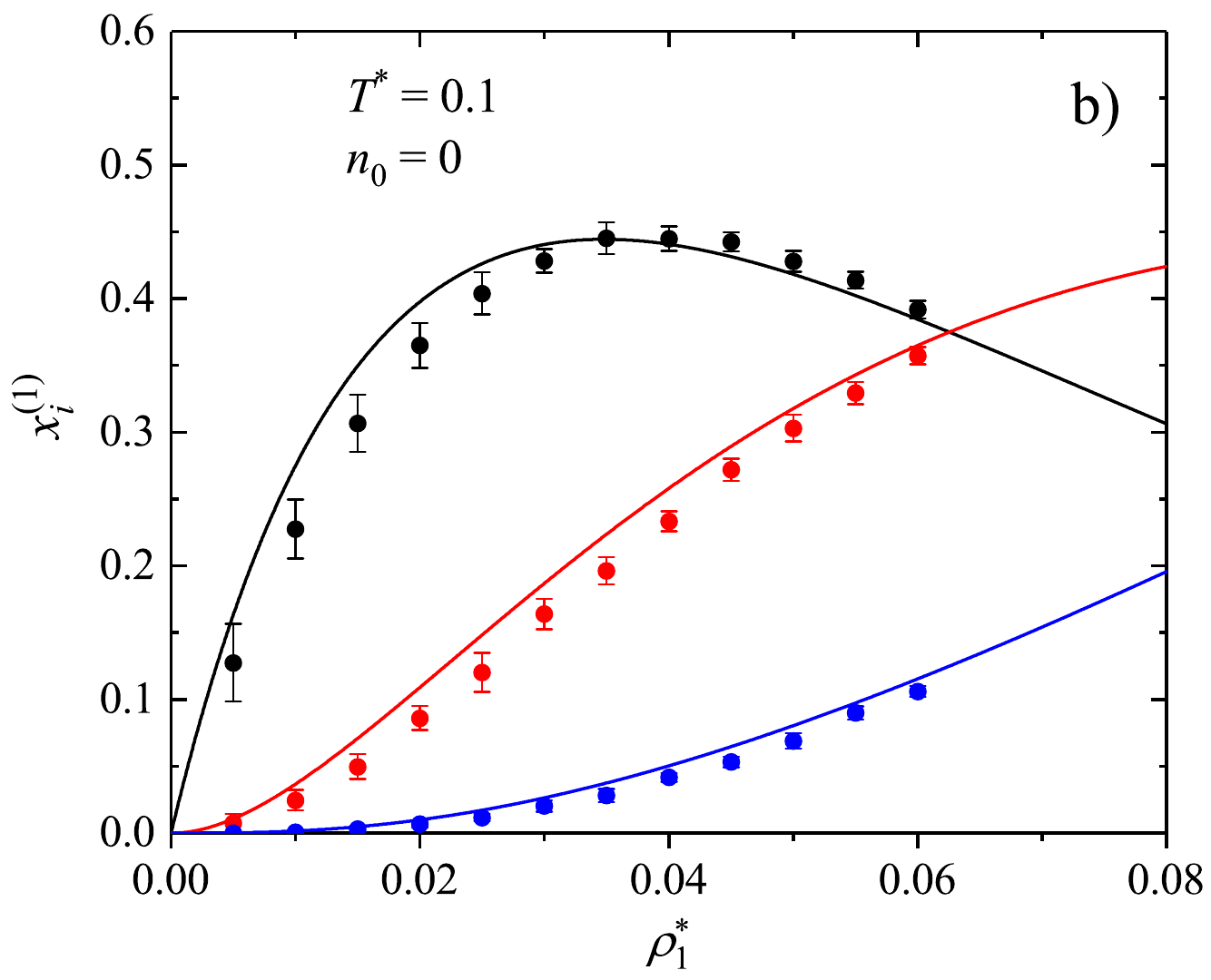}        
		\caption{(Colour online) Fractions of $i$-times bonded particles \textit{vs} density $\rho^*_1$ for the system with $\epsilon^*_{11}=1$, $n_1=3$, $n_0=0$, $\eta_0=0.1$, $\sigma_0/\sigma_1=3$, at temperatures $T^*=0.085$ (panel a) and $T^*=0.1$ (panel b). Here, $x_1^{(1)}$, $x_2^{(1)}$, and $x_3^{(1)}$ are shown by black, red, and blue lines, respectively. Symbols represent simulation results.}
		\label{f1}
	\end{center}
\end{figure}

The equations of motion were integrated using the velocity Verlet algorithm with a timestep of $\delta t = 0.0001\tau$. 
The Langevin damping time was set to $\tau_d = 0.1\tau$, where $\tau = \left(m\sigma^2/\epsilon\right)^{1/2}$ defines the unit of time.
The relatively small timestep was required to ensure a numerical stability of the binding dynamics between antibody molecules and obstacles. 
In preliminary simulations, larger timesteps were found to destabilize the bonded configurations due to the steep repulsive interactions and the very short attractive range $r_w$ employed in the model.

All simulations were carried out in a cubic box of fixed size $L = 30\sigma$ with periodic boundary conditions applied in all three spatial directions. 
The number of obstacle particles was determined from the obstacle packing fraction $\eta_0$ according to $N_0 = 6 \eta_0 L^3/(\piup \sigma_0^3)$,
which yields $N_0\approx 191$ for $\eta_0 = 0.1$, when the obstacle diameter $\sigma_0 = 3\sigma$.
The number density of antibody molecules was varied in the range $\rho_1 = 0.005-0.06\sigma^{-3}$. For Y-shaped molecules composed of seven beads 
of diameter $\sigma_1 = \sigma$, this corresponds to a maximum molecular packing fraction $\eta_1 = 7\piup \sigma_1^3 \rho_1/6 \approx 0.22$ 
at the highest density.

While the molecular architecture of the antibodies remained unchanged throughout the simulations, different numbers of attractive patches on the obstacle particles were considered, specifically $n_0 = 0$, $1$ and $3$. The energetic parameters $\epsilon_{ij}$ in the simulations were set equivalent for molecule-molecule and molecule-obstacle interactions.
The systems were investigated at two temperatures, $T^* = 0.085$ and $0.1$. At these temperatures, the average numbers of bonds formed by patches between molecules and between molecules and obstacles were calculated at the different densities $\rho_1$. 
A bond between two patches was considered to be formed when the distance between them satisfied $r \leqslant r_w$, i.e., when it was within the attractive well range of the CSW potential.

For each set of parameters, the simulations were carried out in three successive stages.
At the first stage, a random configuration of non-overlapping obstacles was setup within the simulation box. Afterwards, the $N_1$ antibody molecules were inserted into the void space not occupied by the obstacles. To eliminate the memory of the initial configuration, a short equilibration run of 1M steps was performed at temperature $T^* = 1$, with the patch-patch interactions switched off.
At the second stage, the patch interactions were activated and the system was simulated at $T^* = 0.085$ or $0.1$ until equilibrium was reached. 
Equilibration process was monitored through the saturation of the average number of bonds formed between patches. 
Depending on the system parameters, this stage required up to 1000M integration steps.
The third stage is the production run, during which statistics on the bonds formed between patches were collected.
Since the bonding statistics depend strongly on the number of molecules, particularly at lower temperature, different production lengths were employed. 
For densities $\rho_1 \leqslant 0.02$, the production run consisted of 1000M steps, whereas for higher densities it was reduced to 200M steps. The numbers of molecules forming one, two and three bonds 
with other molecules and with obstacle particles were sampled every 10K steps and subsequently averaged over the production run in order to make a comparison with the predictions of the theory presented in our study.
{Error bars were calculated as the standard deviation of the sampled data.}

\begin{figure}[!htb] 	
	\begin{center}
		\includegraphics[width = 0.45\textwidth]{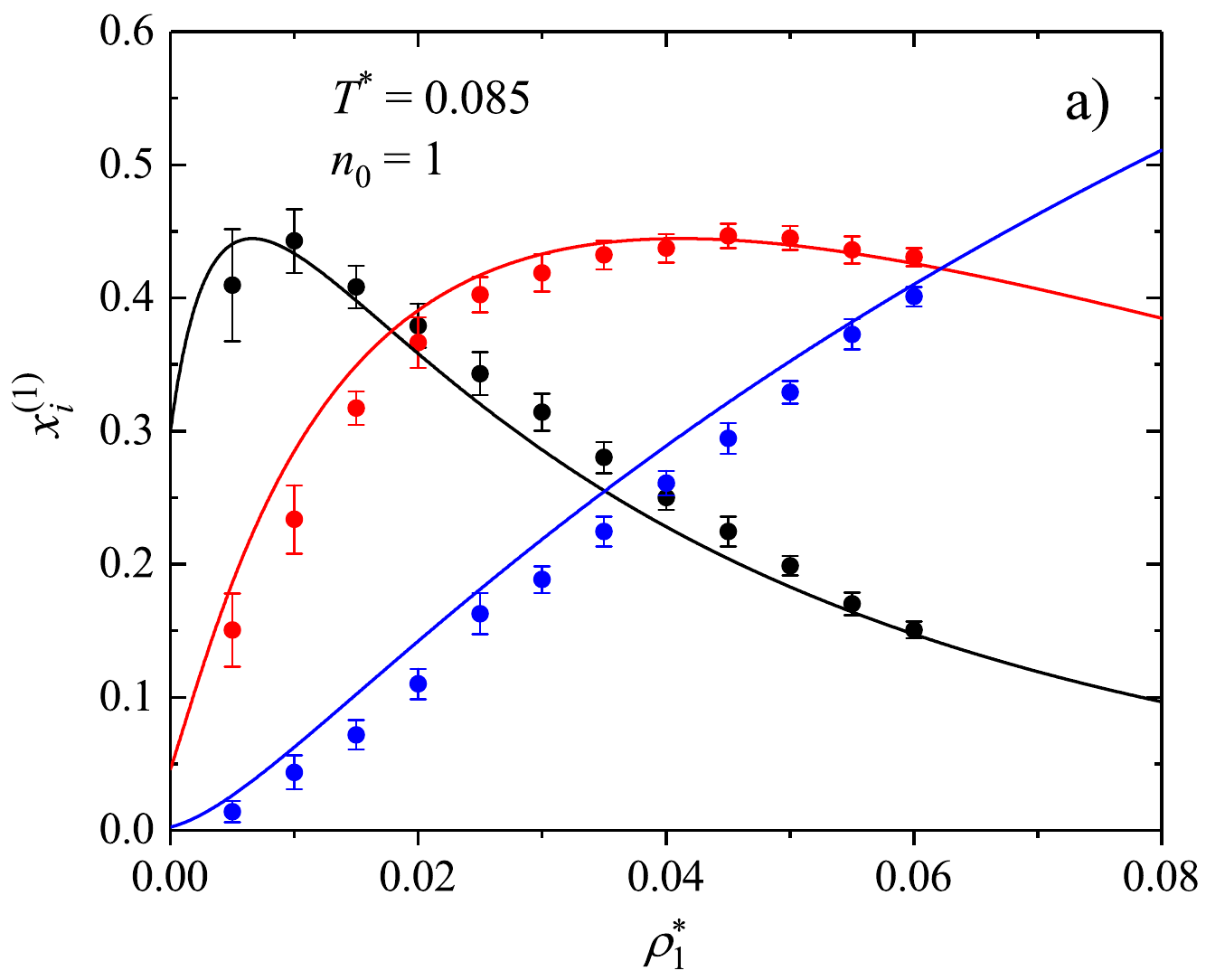}
		\includegraphics[width = 0.45\textwidth]{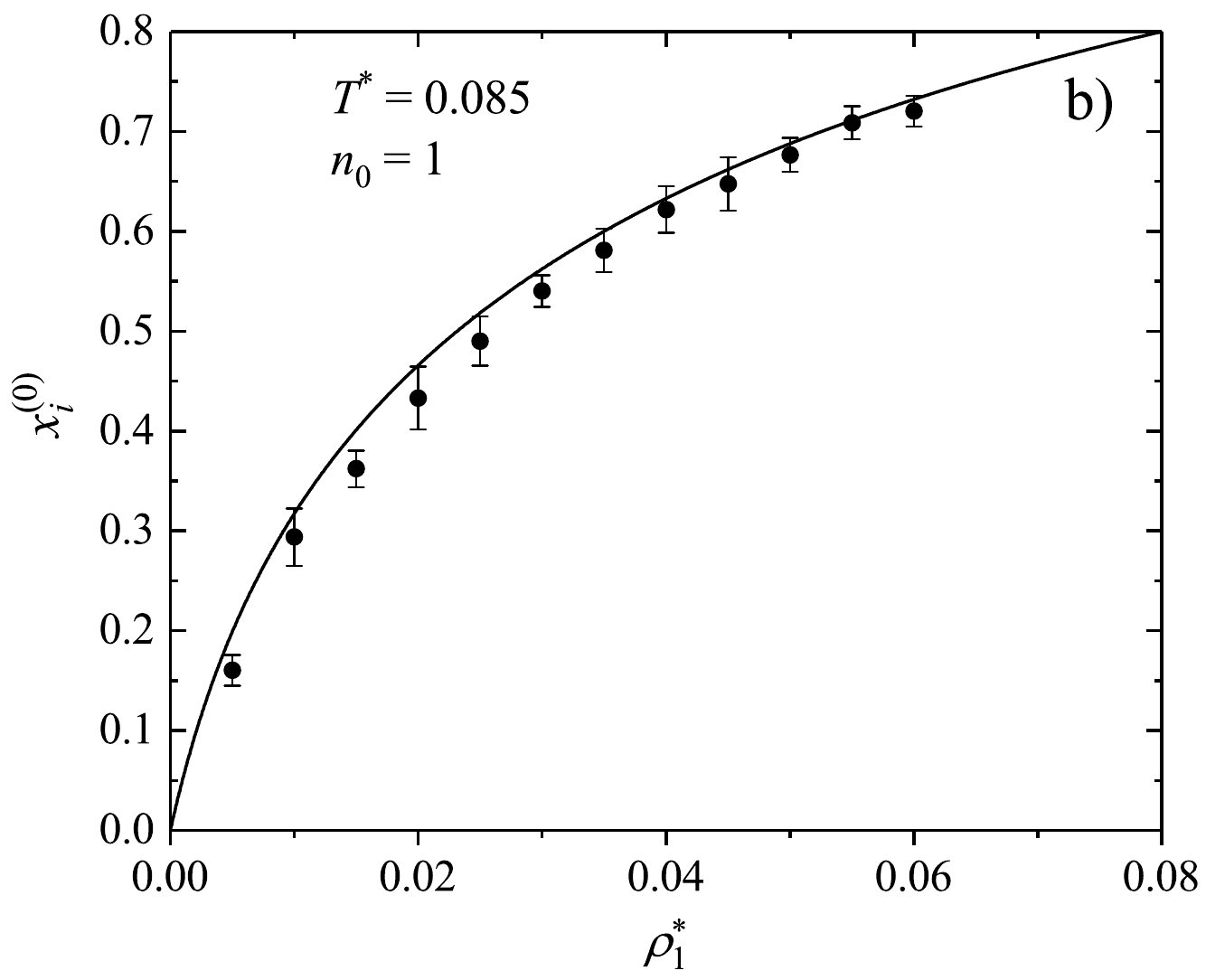}
		\caption{(Colour online) Fractions of $i$-times bonded molecules \textit{vs} density $\rho^*_1$ for the system with $\epsilon^*_{11}=1$, $\epsilon^*_{01}=1$, $n_1=3$, $n_0=1$, $\eta_0=0.1$, $\sigma_0/\sigma_1=3$, at temperature $T^*=0.085$. In panel~a, $x_1^{(1)}$, $x_2^{(1)}$, and $x_3^{(1)}$ are shown by black, red, 
			and blue lines, respectively, while in panel~b $x_1^{(0)}$ is shown by a black line.
			Symbols represent simulation results.}
		\label{f2}
	\end{center}
\end{figure}

\begin{figure}[!htb] 	
	\begin{center}
		\includegraphics[width = 0.45\textwidth]{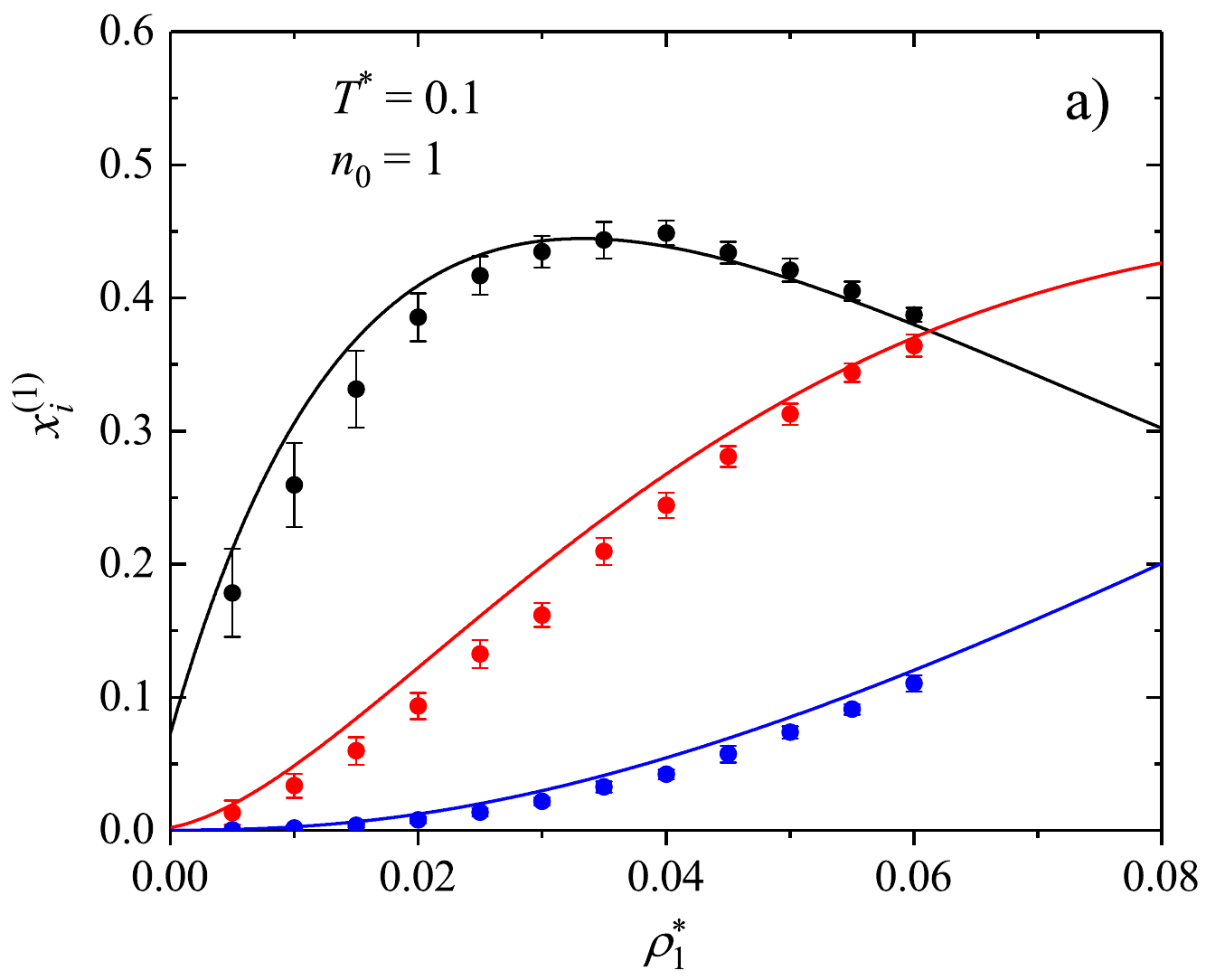}
		\includegraphics[width = 0.45\textwidth]{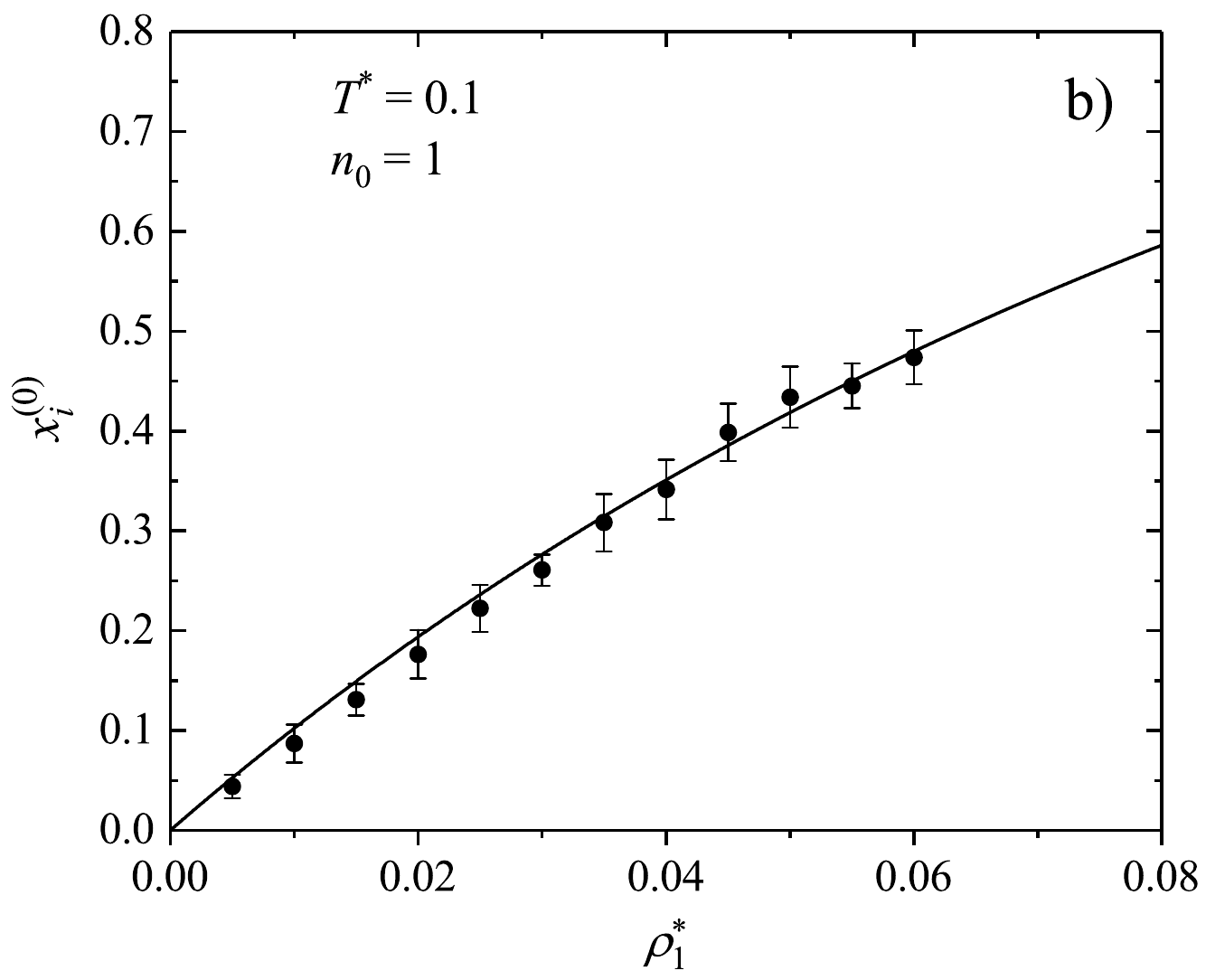}
		\caption{(Colour online) The same as in figure~\ref{f2}, but at temperature $T^*=0.1$.}
		\label{f2-2}
	\end{center}
\end{figure}

\section{Results and discussion}

\begin{figure}[!htb] 	
	\begin{center}
		\includegraphics[width = 0.45\textwidth]{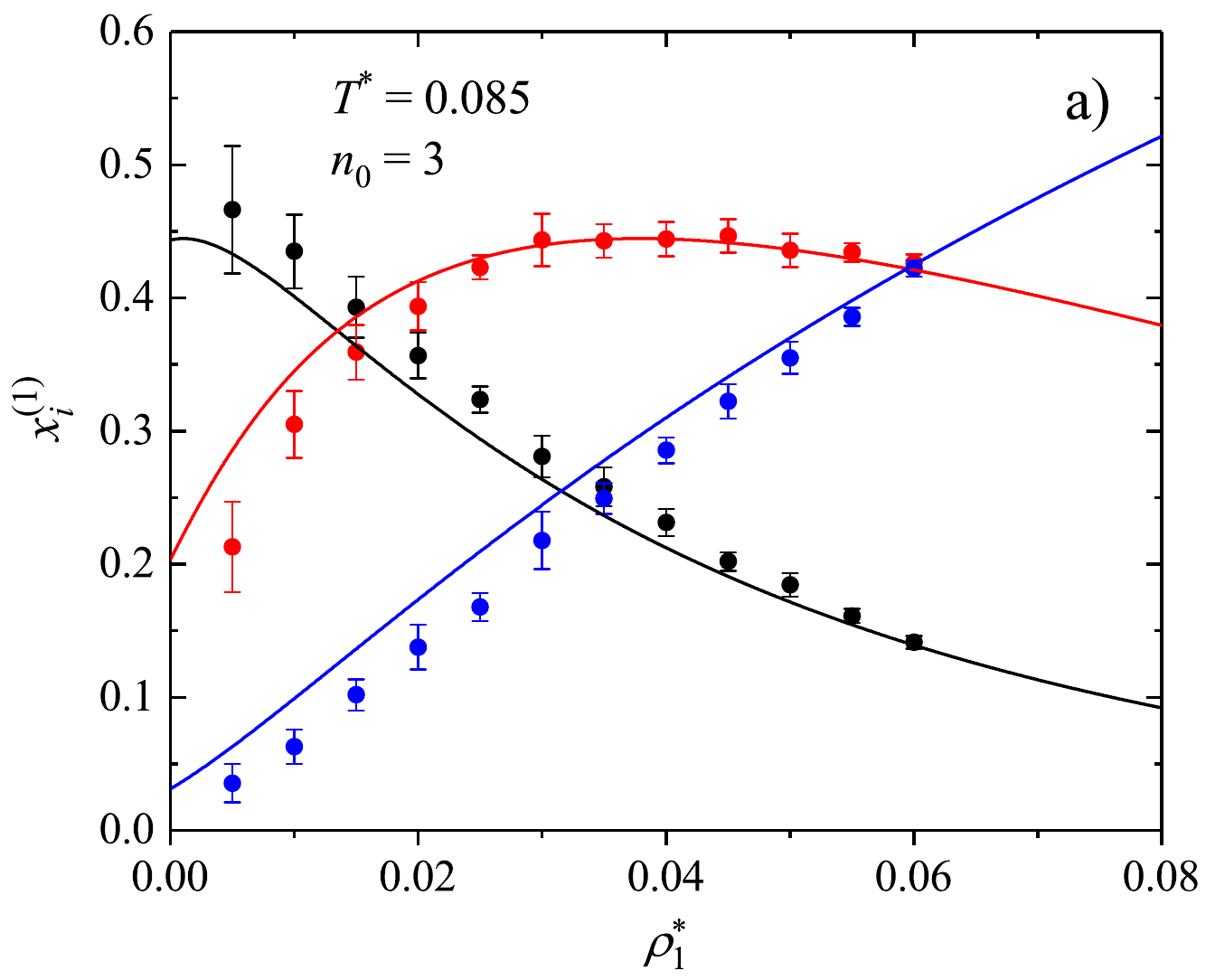}
		\includegraphics[width = 0.45\textwidth]{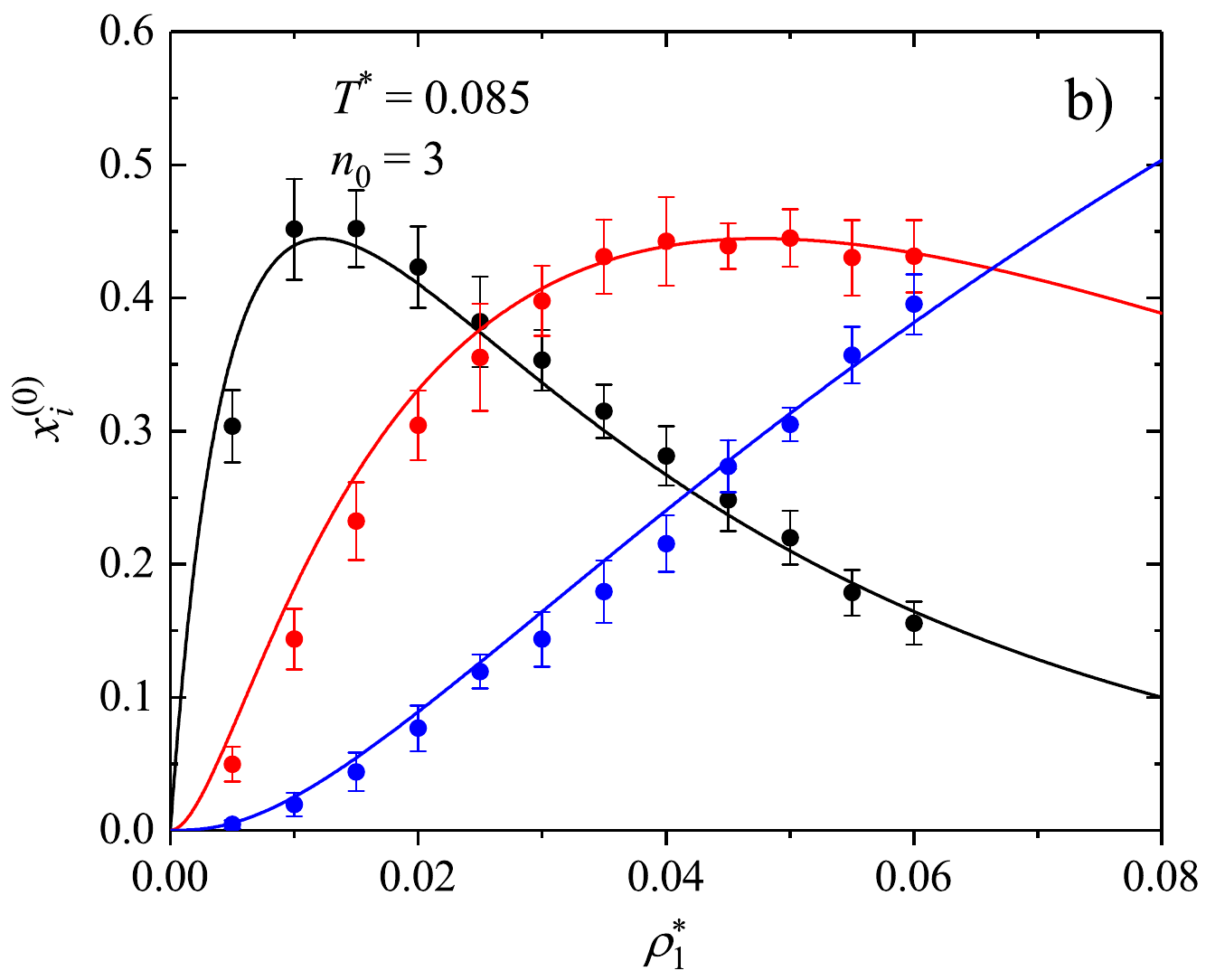}
		\caption{(Colour online) Fractions of $i$-times bonded molecules \textit{vs} density $\rho^*_1$ for the system with $\epsilon^*_{11}=1$, $\epsilon^*_{01}=1$, $n_1=3$, $n_0=3$, $\eta_0=0.1$, $\sigma_0/\sigma_1=3$, at temperature $T^*=0.085$. In panel~a, $x_1^{(1)}$, $x_2^{(1)}$, and $x_3^{(1)}$ are shown by black, red, and blue lines, respectively, and in panel~b $x_1^{(0)}$, $x_2^{(0)}$, and $x_3^{(0)}$ are shown by black, red, and blue lines, respectively.
		Symbols represent simulation results.}
		\label{f3}
	\end{center}
\end{figure}

To assess the accuracy of the theory we calculate the fractions of $i$-times bonded molecules $x_i$ as a 
function of their density $\rho^*_1=\rho_1/\sigma^3$, when the matrix obstacles packing fraction $\eta_0=0.1$ and 
the number of patches placed on the matrix obstacles $n_0=0$, $1$ and $3$. 
The value $n_0=0$ means that the matrix obstacles have no patches.
The strength of attractive interaction between patches of the molecules and between the molecules and the matrix obstacles is set to 
$\epsilon^*_{11}=\epsilon_{11}/\epsilon=1.0$ and $\epsilon^*_{01}=\epsilon_{01}/\epsilon=1.0$, respectively. 
These fractions are the key quantities of the 
theory and to a substantial degree their accuracy define the accuracy of the final results.  
Numerical calculations were performed for the $\sigma_1=\sigma$ and $\sigma_0=3\sigma$, and the temperatures $T^*=0.085$ and $T^*=0.1$.
In figures~\ref{f1}--\ref{f3}, we present the results for the fractions $x_i^{(1)}$ and $x_i^{(0)}$ of $i$-times bonded molecules. We note that the superscript~$(0)$ refers to the bonds formed only between molecules and obstacles, while the superscript $(1)$ denotes both molecule-obstacle and molecule-molecule bonds.
Overall, a reasonably good quantitative agreement between the theoretical predictions and the simulation results is observed, especially at the higher temperature $T^*=0.1$. 
However, at the temperature $T^*=0.085$ and higher densities, the theory becomes less accurate.
In particular, at lower densities it tends to overestimate the fractions $x_i^{(1)}$ and $x_i^{(0)}$ for $n_0 = 0$ and $1$, and surprisingly underestimates $x_1^{(1)}$ and $x_1^{(0)}$ for $n_0 = 3$.
A comparison of the results in figures~\ref{f1}--\ref{f3} at different temperatures shows that at $T^* = 0.1$ the presence of patches on the matrix has only a weak effect on the bonding fractions; although as the temperature decreases, this effect becomes more pronounced.

\begin{figure}[!htb] 	
	\begin{center}
		\includegraphics[width = 0.45\textwidth]{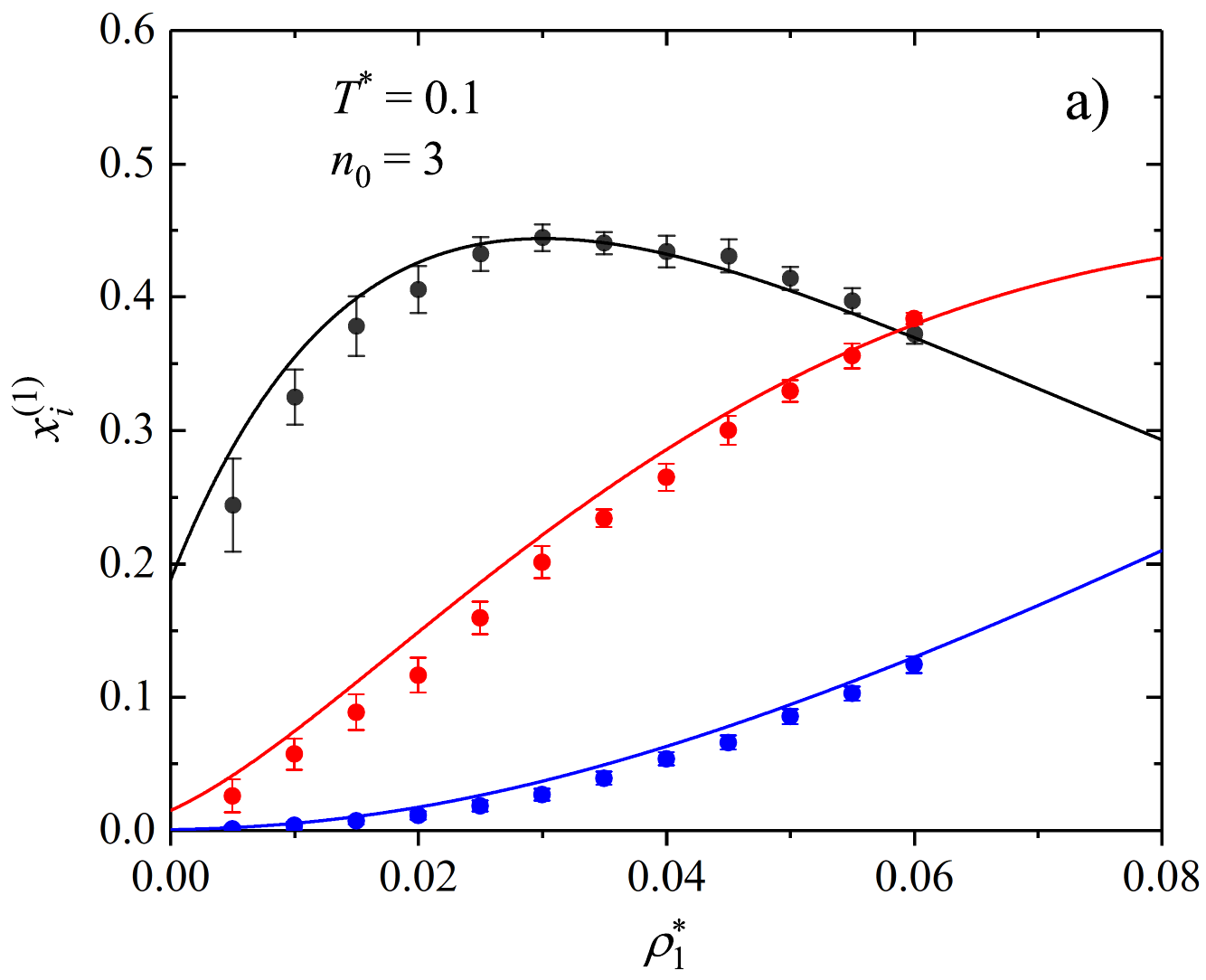}
		\includegraphics[width = 0.45\textwidth]{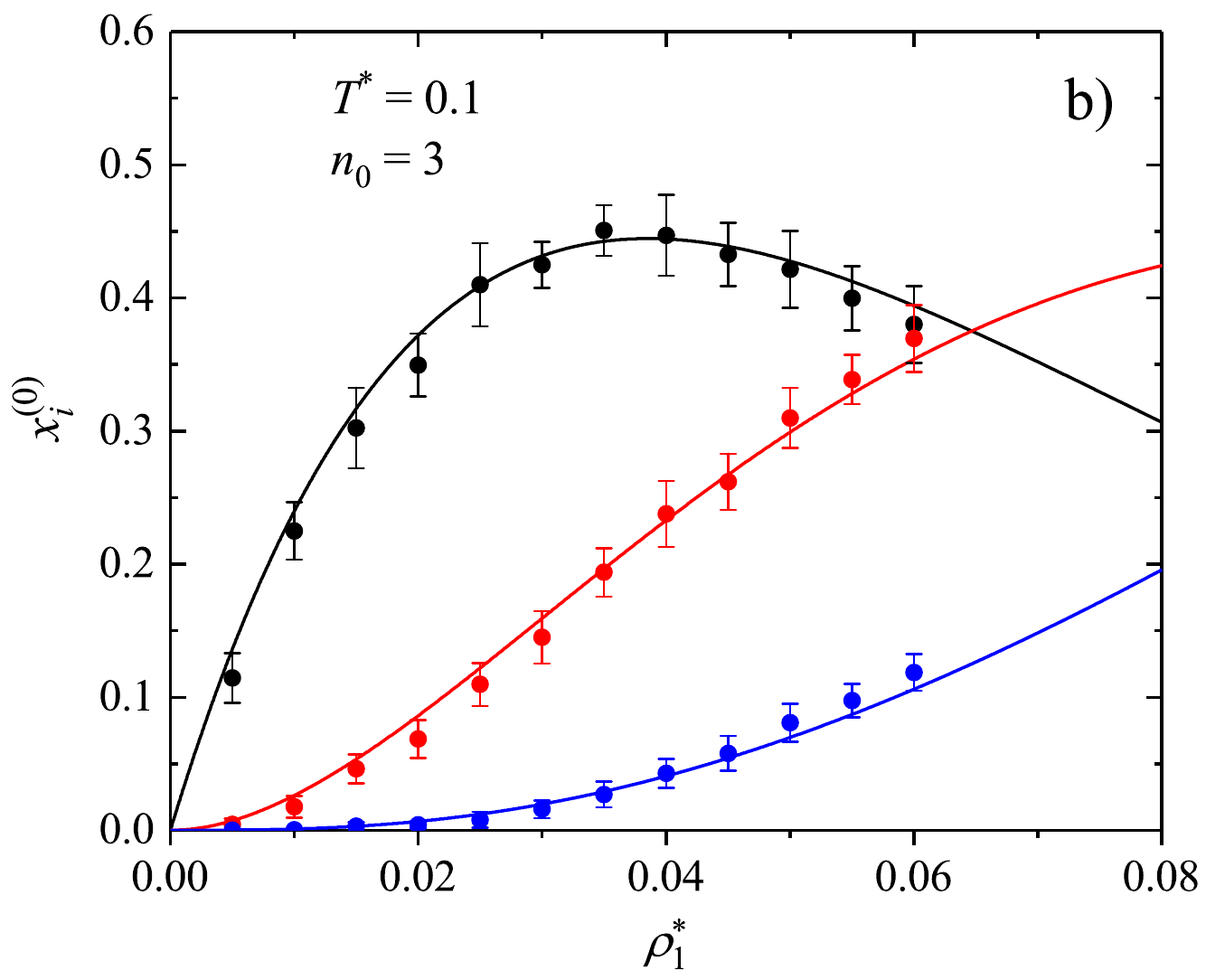}
		\caption{(Colour online) The same as in figure~\ref{f3}, but at temperature $T^*=0.1$.}
		\label{f3-1}
	\end{center}
\end{figure}

\begin{figure}[!htb] 	
	\begin{center}
		\includegraphics[width = 0.6\textwidth]{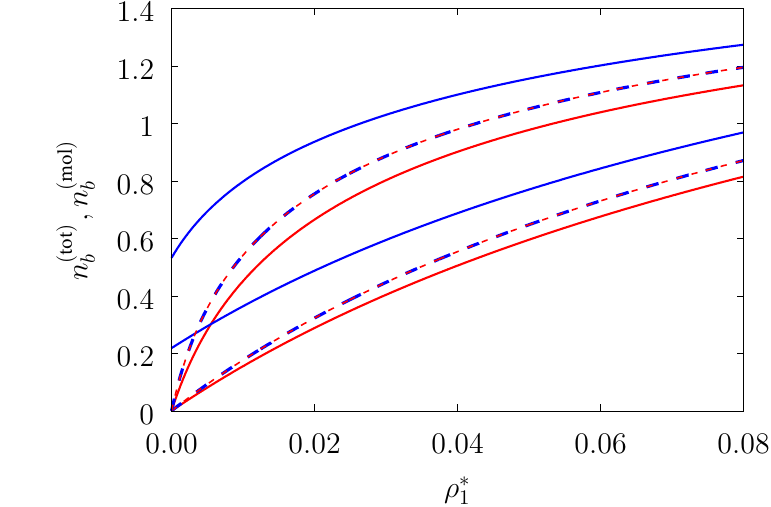}
		\caption{(Colour online) Number of bonds per antibody molecule \textit{vs} density $\rho_1^*$ at $T^*=0.85$ (upper series of the curves) and $T^*=0.1$ (lower series of the curves) and for the model with $n_0=0$ (dashed lines) and 
        $n_0=3$ (solid lines). Here red lines denote $n_b^{\rm{(mol)}}$ and blue lines denote $n_b^{\rm{(tot)}}$. The rest of the model parameters are:
			$n_1=3$, $\epsilon_{11}^*=1$, $\epsilon_{01}^*=1$ and $\eta_0=0.1$.}
		\label{f4}
	\end{center}
\end{figure}

In all cases, the fraction of singly bonded molecules initially increases with increasing density and, after reaching a maximum at a certain density, decreases. A similar behavior is observed for the fraction of doubly bonded molecules, although the position of the corresponding maximum is shifted toward higher densities. By contrast, the fraction of triply bonded molecules increases monotonously with increasing density.
This behavior reflects the general tendency of the system to form a three-dimensional network of bonds as the density increases, through the gradual formation of clusters of increasing size. In the presence of patchy matrix obstacles, this process results from the competition between the bond formation among molecules and bonding between molecules and matrix obstacles. 
In figure~\ref{f4}, we present the total number of bonds per molecule, $n_b^{\rm(tot)}$, along with the number of bonds connecting only the molecules normalized by the number of molecules, $n_b^{\rm{(mol)}}$, 
as functions of density at $T^*=0.085$ and $T^*=0.1$ for the models with $\eta_0=0.1$ and $n_0=0,\,3$. 
These quantities were calculated using the following expressions:
\be
n^{\rm{(tot)}}_b={1\over 2}\sum_{i=1}^{n_1}ix_i^{(1)}+{1\over 2}{\rho_0\over\rho_1}
\sum_{i=1}^{n_0}ix_i^{(0)},
\label{xbt}
\ee
\be
n^{\rm{(mol)}}_b={1\over 2}\sum_{i=1}^{n_1}ix_i^{(1)}-{1\over 2}{\rho_0\over\rho_1}
\sum_{i=1}^{n_0}ix_i^{(0)}.
\label{xbl}
\ee
%
At low densities, $n_b^{\rm{(mol)}}$ is nearly identical for models with and without patches on
the matrix obstacles and most of the particles of the liquid are bonded to the matrix
obstacles. As the density increases, $n_b^{\rm{(mol)}}$ for the model with a patchy matrix
becomes lower than that for the model without patches on the matrix obstacles. Thus, at
higher densities, the formation of a network of bonds connecting fluid particles is delayed
due to the presence of the patchy matrix. This effect is more pronounced for the lower temperature.

\begin{figure}[!htb] 	
	\begin{center}
		\includegraphics[width = 0.6\textwidth]{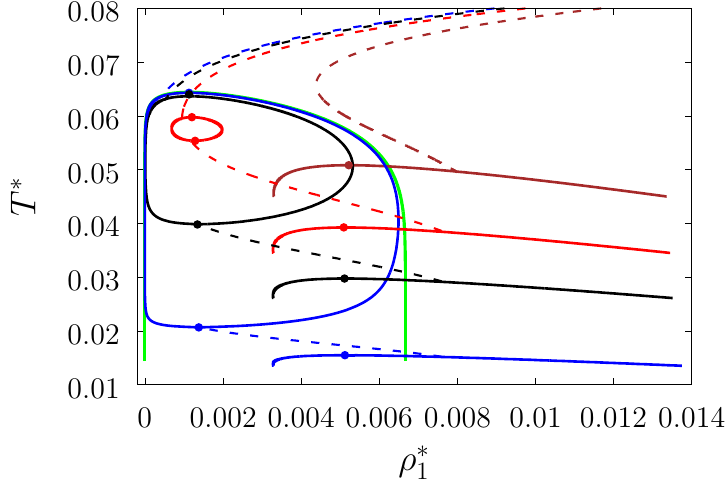}
		\caption{(Colour online) Liquid-gas phase diagram (solid lines) and percolation thresholds (dashed lines) for the model with $n_0=0$ (green line) and $n_0=1$, $\omega_{01}=0.1\sigma_1$ and
			$\epsilon^*_{01}=0.63$ (blue lines), $\epsilon^*_{01}=0.75$ (black lines), 
			$\epsilon^*_{01}=0.83$ (red lines) and $\epsilon^*_{01}=0.97$ (brown lines).
			Here circles denote position of the corresponding critical points.}
		\label{f5}
	\end{center}
\end{figure}

Finally in figure~\ref{f5} we present the phase behavior of the model with $n_0=1$,
$\omega=0.1\sigma_1$, $\eta_0=0.2$ at $\epsilon^*_{01}=0.63,\;0.75,\;0.83,\;0.97$.
In the same figure, we also show the corresponding percolation threshold lines, which separate percolating and non-percolating regions. In the non-percolating region, antibody molecules form finite-size clusters, whereas in the percolating region they form an infinite bonded network.

Owing to the competition between the bond formation among molecules and bonding between molecules and matrix obstacles, the resulting phase behavior is highly nontrivial.
For $n_0=0$, the phase diagram has the usual shape and exhibits a single critical point.  
However as $\epsilon_{01}$ increases, the phase diagram splits into two distinct regions, giving rise to three critical points.
At higher temperatures, the phase behavior is characterized by a closed-loop coexistence region with both upper and lower critical points, while at lower temperatures it reduces to the conventional liquid-gas coexistence of a one-component fluid with a single critical point. These two regimes are separated by an intermediate temperature interval in which no phase coexistence occurs, clearly demonstrating the re-entrant nature of the phase behavior.
At higher temperatures, phase separation is driven by the formation of a bonded network connecting the molecules, with molecules-matrix bonding having only a minor effect. As the temperature decreases, the bonding between molecules and matrix obstacles becomes increasingly significant and eventually disrupts the fluid-fluid network, thereby suppressing the phase separation. Upon further cooling, all matrix patches become saturated, allowing the fluid-fluid network to reform and restoring the phase separation. The size of the closed-loop coexistence region depends on the strength of the molecule-obstacle interaction~$\epsilon^*_{01}$: with increasing $\epsilon^*_{01}$, this region shrinks and eventually disappears for $\epsilon_{01}^*> 0.93$. At the same time, the low-temperature part of the phase diagram shifts toward higher temperatures.

\section{Conclusions}

In this work, we have presented a theoretical study of a simple model of monoclonal antibodies confined in a patchy random porous medium. Antibody macromolecules are modelled as Y-shaped assemblies of tangentially connected seven hard-sphere beads, with three terminal beads bearing sticky patches, while the porous matrix is represented by randomly distributed hard-sphere obstacles that are also decorated with sticky patches. The model is designed to capture the essential features of antibody behavior in crowded biological environments and to investigate the role of strong short-range attractive interactions, so-called soft effects between molecules and matrix obstacles.
The theoretical description was developed by combining Wertheim’s multidensity thermodynamic perturbation theory with the Flory-Stockmayer theory of polymerization and scaled particle theory for a fluid in porous media. Within this framework, we analyzed the thermodynamic, percolation, and phase behavior of the system. The predictions of the theory were validated against computer simulation data. Overall, a reasonably good quantitative agreement was obtained for the fractions of $i$-times bonded molecules, demonstrating the reliability of the proposed approach.
Our results show that the properties of the system are governed by the competition between attractive interactions among molecules and those between molecules and matrix obstacles. This competition gives rise to a highly nontrivial behavior, including a re-entrant phase separation. At higher temperatures, the phase separation is associated with the formation of a percolating network of bonds between antibody molecules and is characterized by a closed-loop coexistence region with both upper and lower critical points. Upon lowering the temperature, the bonding between molecules and matrix obstacles becomes dominant, leading to the disruption of the molecular network and suppression of phase separation. At still lower temperatures, saturation of the matrix patches restores the fluid-fluid network and the conventional liquid-gas coexistence behavior.

The present study highlights the importance of confinement and matrix-induced bonding effects in determining the collective behavior of associating macromolecular systems. The proposed model and theoretical framework provide a useful basis for further investigations of antibody solutions and other complex associating fluids in crowded and heterogeneous environments.

\section{Funding}

Yu.V.K. acknowledges financial support through the MSCA4Ukraine project (ID: 101101923), funded by the
European Union. 

\bigskip
\bibliographystyle{cmpj}
\bibliography{tadeja}

\ukrainianpart

\title{Властивості флюїду модельних антитіл, обмеженого жорсткими сферичними перешкодами: ефекти зв’язування між перешкодами та антитілами}
\author{Ю. В. Калюжний\refaddr{label1}, Т. Пацаган\refaddr{label2,label3}}
\addresses{
	\addr{label1} Факультет хiмiї та хiмiчної технологiй унiверситету Любляни, Вечна пот 113, SI-1000, Любляна, Словенiя
	\addr{label2} Інститут фізики конденсованих систем
	імені І. Р. Юхновського НАН України, 79011  Львів, вул.~Свєнціцького, 1, Україна 
	\addr{label3} Інститут прикладної математики та фундаментальних наук, Національний університет ``Львівська політехніка'', вул. С. Бандери 12, 79013 Львів, Україна
}

\makeukrtitle

\begin{abstract}
	\tolerance=3000%
	Ми досліджуємо спрощену модель моноклональних антитіл, обмежених у випадковому пористому середовищі з плямистою (patchy) структурою. Антитіла подано у вигляді Y-подібних частинок, що складаються із семи дотичних твердих сфер з притягальними сайтами на кінцевих мономерах, тоді як матриця складається з випадково розподілених перешкод у вигляді твердих сфер, які містять адгезійні сайти. Модель відтворює поведінку антитіл у скупчених біологічних середовищах за наявності сильних короткодіючих притягальних взаємодій антитіло–матриця.
	Теоретичний підхід поєднує багатогустинну термодинамічну теорію збурень Вертгайма, теорію полімеризації Флорі–Стокмаєра та теорію масштабованої частинки для флюїдів у пористих середовищах. Ми аналізуємо термодинамічні властивості, пороги перколяції та фазову поведінку, а також порівнюємо окремі результати з новими комп’ютерними симуляціями.
	Взаємодія між взаємодіями антитіло–антитіло та антитіло–матриця зумовлює складну фазову поведінку, включаючи реентрантне фазове розділення із замкненою областю співіснування за вищих температур і звичайне розділення типу рідина–газ за нижчих температур.

	\keywords моноклональні антитіла, макромолекулярне скупчення, модель частинок із локалізованими сайтами взаємодії (плямисті частинки), термодинамічна теорія збурень, перколяція, фазове розділення
\end{abstract}

\end{document}